\documentclass[sigconf,screen]{acmart}

\usepackage{booktabs}
\usepackage{amsmath}
\usepackage{graphicx}
\usepackage{subcaption}
\usepackage{algorithm}
\usepackage{algorithmicx}
\usepackage[noend]{algpseudocode} 
\usepackage{float}                 
\usepackage{enumitem}
\usepackage{balance}
\usepackage{xspace}
\usepackage{multirow}
\usepackage{microtype}
\usepackage{mathtools}
\usepackage{soul}
\sethlcolor{red!20}
\usepackage{hyperref}
\usepackage{tabularx}

\newcommand{\rqbox}[1]{%
  \par\noindent
  \begingroup
    \setlength{\fboxrule}{0pt}
    \setlength{\fboxsep}{2pt}
    \colorbox{gray!10}{%
      \parbox{\dimexpr\linewidth-2\fboxsep\relax}{%
        \vspace{0.5pt}
        #1%
      }%
    }%
  \endgroup
  \par
}

\newboolean{showcomments}
\setboolean{showcomments}{true}
\ifthenelse{\boolean{showcomments}}
{
	\definecolor{myyellow}{RGB}{255, 228, 26}
	\definecolor{myblue}{RGB}{50, 50, 220}
	\newcommand{\nb}[2]{
		{\sf
			\fcolorbox{myyellow}{yellow}{\scriptsize\textbf{#1}}%
			$\blacktriangleright$%
			{\color{myblue}\fontsize{7pt}{8pt}\selectfont\textbf{#2}}%
		}%
	}
}
{
	\newcommand{\nb}[2]{}
}

\newcommand{\codeguardian}{\textsc{CodeGuardian}\xspace} %
\newcommand{\head}[1]{\noindent\textbf{#1.}}

\begin{document}

\title{Large Language Models for Secure Code Assessment: A Multi-Language Empirical Study}

\author{Kohei Dozono}
\orcid{0009-0009-2280-0707}
\email{kohei.dozono@tum.de}
\authornotemark[1]
\affiliation{%
  \institution{Technical University of Munich}
  \country{Germany}
  \postcode{85748}
}

\author{Tiago E. Gasiba}
\orcid{0000-0003-1462-6701}
\email{tiago.gasiba@siemens.com}
\authornotemark[2]
\affiliation{%
  \institution{Siemens AG}
  \country{Germany}
}

\author{Andrea Stocco}
\orcid{0000-0001-8956-3894}
\authornotemark[1]
\email{andrea.stocco@tum.de}
\affiliation{%
  \institution{Technical University of Munich}
  \country{Germany}
}

\renewcommand{\shortauthors}{Dozono et al.}


\begin{abstract}
Most vulnerability detection studies rely on datasets tied to specific programming languages and within-project settings, limiting language diversity and the assessment of model generalizability. Consequently, the effectiveness of deep learning methods, including large language models (LLMs), beyond such restricted scenarios remains largely unexplored.
In this paper, we evaluate six state-of-the-art LLMs (GPT-3.5-Turbo, GPT-4 Turbo, GPT-4o, CodeLlama-7B, CodeLlama-13B, Gemini 1.5 Pro) on vulnerability detection and CWE classification across five languages (Python, C, C++, Java, JavaScript). We compiled a multi-language dataset from diverse sources and investigated different prompt and role strategies. Our results show that GPT-4o achieves the best performance, particularly in few-shot settings.
To examine practical applicability, we developed \codeguardian, a VSCode extension that enables real-time LLM-assisted vulnerability detection. In a user study with 22 professional developers, \codeguardian doubled accuracy and halved task completion time compared to a control group.
\end{abstract}

\copyrightyear{2026}
\acmYear{2026}
\setcopyright{cc}
\setcctype{by}
\acmConference[DeepTest '26]{7th International Workshop on Deep Learning for Testing and Testing for Deep Learning (DeepTest '26)}{April 12--18, 2026}{Rio de Janeiro, Brazil}
\acmBooktitle{7th International Workshop on Deep Learning for Testing and Testing for Deep Learning (DeepTest '26) (DeepTest '26), April 12--18, 2026, Rio de Janeiro, Brazil}
\acmPrice{}
\acmDOI{10.1145/3786154.3788584}
\acmISBN{979-8-4007-2386-5/2026/04}



\maketitle

\section{Introduction}\label{sec:introduction}

Software vulnerabilities are defects that can enable attackers to steal data, gain control of systems, or insert malware~\cite{shirey2000internet,shoshitaishvili2015firmalice,10.1145/3329786}. They arise from design flaws, insecure code, or poor data and access management. In this work, we focus on vulnerabilities rooted in \textit{source code bugs}. According to the U.S. Department of Homeland Security, 90\% of security incidents exploit such flaws~\cite{cisa_software_assurance_2021}. Yet, over half of developers report difficulties identifying vulnerabilities~\cite{gitlab_devsecops_survey_2022}, highlighting the need for effective detection tools.

Traditional program-analysis techniques~\cite{5386616,9700400,Thapa2022,Alqarni2022Low,Chen2023,6606613,Rangnau2020} are limited by high false positives, narrow coverage, and long analysis times~\cite{6606613,Rangnau2020}. Deep learning approaches~\cite{Russel2018,HANIF2021103009,Dam2021,2020-Humbatova-ICSE} improved scalability but often treat code as text, failing to capture semantic structure~\cite{2020-Humbatova-ICSE,2020-Riccio-EMSE}. Large Language Models (LLMs), pre-trained on massive code and text corpora~\cite{NIPS2017_3f5ee243}, have recently shown promise for vulnerability detection~\cite{Thapa2022,Alqarni2022Low,Chen2023}. However, existing evaluations are limited to one or two languages (e.g., BigVul~\cite{10.1145/3379597.3387501}, SVEN~\cite{10.1145/3576915.3623175}), and LLM capabilities across diverse languages remain largely unexplored. Moreover, most studies remain purely quantitative, with little evidence on \textit{how LLMs can assist developers in practice}.

This paper addresses these gaps by evaluating six state-of-the-art LLMs (GPT-3.5-Turbo, GPT-4 Turbo, GPT-4o, CodeLlama-7B, CodeLlama-13B, Gemini 1.5 Pro) on vulnerability detection and CWE classification across five languages (Python, C, C++, Java, JavaScript). We focus on the top 25 CWE classes~\cite{MITRE2023CWE} and compile a dataset of over 370 manually validated vulnerable snippets. Beyond our quantitative analysis, we investigate practical applicability through \codeguardian, a VSCode extension that integrates LLMs into developers' workflows. In a user study with 22 industry practitioners, \codeguardian doubled accuracy in CWE classification and reduced task completion time by 60\%.
The contributions of this work are as follows:

\noindent
\head{Dataset} A curated dataset of 370+ vulnerabilities across five languages, supporting evaluation of detection and classification tools.

\noindent
\head{Evaluation} An empirical study of six LLMs on vulnerability detection and classification across multiple languages. 

\noindent
\head{\codeguardian} A VSCode extension enabling real-time vulnerability scanning, released in our replication package~\cite{replication-package}.

\noindent
\head{User Study} A qualitative evaluation with 22 professional developers, showing significant improvements in speed and accuracy with \codeguardian.

\section{Empirical Study}\label{sec:approach}

\subsection{Research Questions}

\noindent
Our study considers the following research questions:

\textbf{RQ\textsubscript{1} (vulnerability detection):}
How effective are LLMs in detecting CWEs across multiple languages? How does effectiveness vary among different languages?
    


\textbf{RQ\textsubscript{2} (CWE classification):}
How effective are LLMs in classifying CWEs across multiple languages? How does effectiveness vary among different languages?

\subsection{Dataset Design} 

The representativeness of datasets is crucial for effective vulnerability detection and classification~\cite{ding2024,Croft2023}.  
Most existing works adopted datasets with limited language diversity, focusing primarily on C/C++~\cite{10.1145/3639476.3639762,10.1145/3611643.3616262,10.1145/3379597.3387501,10.1145/3576915.3623175,10.5555/3454287.3455202,10.1109/ICSE-SEIP52600.2021.00020}.
Although some researchers have attempted to address language diversity limitations with evaluating LLMs on three to seven programming languages~\cite{Shu2025,Atiiq2024,Zhang2025}, they either focus solely on binary vulnerability detection without CWE classification~\cite{Shu2025,Atiiq2024}. Additionally, most existing studies rely heavily on publicly available vulnerability data primarily from the National Vulnerability Database (NVD), potentially limiting the diversity and real-world applicability of their findings.

In contrast, in this work, we consider software vulnerabilities of five popular programming languages, namely Python, C, C++, Java, and JavaScript, which represent the top-ranked languages according to IEEE's 2025 survey~\cite{ieee_spectrum_top_languages_2025}. Furthermore, we compiled a dataset that integrates three existing vulnerability datasets: CVEFixes, CWE-snippets, and the JavaScript Vulnerability DataSet (JVD). 
\autoref{tab:refds-list} shows the statistics of these datasets, including the number of languages, CWEs, and files. 
CVEFixes is a fix-commit vulnerability dataset~\cite{Bhandari_2021}. 
CWE-snippets is a dataset created in collaboration with an industrial partner for this study. It consists of numerous vulnerability snippets across multiple programming languages. 
JVD is a vulnerability dataset for JavaScript code~\cite{Ferenc2019}.
From each dataset, we selected the top 25 CWE most dangerous weaknesses~\cite{MITRE2023CWE} for each of the five selected programming languages. 

\begin{table}[t]
\def\arraystretch{1.1}%
\setlength{\tabcolsep}{13.5pt}
\caption{Vulnerabilities benchmarks evaluated in this study.}
\centering
\begin{tabular}{@{}lccc@{}}
\toprule
\textbf{Dataset} & \textbf{\# Languages} & \textbf{\# CWEs} & \textbf{\# Files} \\
\midrule
CVEFixes \cite{Bhandari_2021} & 27 & 180 & 18,249 \\
CWE-snippets & 42 & 506 & 231,571 \\
JVD \cite{Ferenc2019} & 1 & N/A & 12,125 \\
\bottomrule
\end{tabular}
\label{tab:refds-list}
\end{table}

\subsubsection{Dataset Construction and Validation}

Given the critical importance of data quality in vulnerability detection research~\cite{ding2024}, we adopted a systematic approach for dataset construction that addresses known limitations in existing vulnerability datasets. Our methodology involves several key phases designed to ensure both representativeness and accuracy of the collected data.

The data collection process involved executing systematic queries against the CVEFixes and CWE-snippets databases to extract code snippets corresponding to specific CWE categories across all target programming languages. Since JVD was structured as a CSV file, we employed systematic filtering techniques to identify relevant JavaScript vulnerabilities. To maintain dataset balance and address the inherent scarcity of certain vulnerability types in specific languages, we set a maximum threshold of three snippets per CWE class across all languages. This constraint was particularly important given that certain vulnerabilities—such as CWE-787 (Out-of-bounds Write)—tend to be associated with memory-unsafe languages like C and C++, while remaining relatively uncommon in memory-safe languages such as Python and Java.

Furthermore, our methodology involved manual validation of all selected code snippets. Each snippet has been inspected to verify both the presence of the vulnerability and the associated CWE ID. This validation process was essential to mitigate the data quality issues that have been identified in existing vulnerability datasets~\cite{ding2024}.

For binary vulnerability detection tasks, we curated non-vulnerable code snippets from multiple sources within our reference datasets. These included fixed versions of vulnerable code from CVEFixes commit histories, solution implementations from the CWE-snippets dataset, and verified secure code samples from JVD. The selection process for non-vulnerable snippets required careful manual inspection to ensure they represented genuinely secure implementations rather than simply alternative vulnerable patterns. It is important to note that the majority of non-vulnerable snippets in our dataset are not direct fixed versions of the corresponding vulnerable snippets. Based on our manual inspection, only a limited number of snippet pairs existed where the non-vulnerable code was the direct fixed version of a specific vulnerable snippet (one pair in C++ and two pairs in C). The use of independent non-vulnerable samples would also support model generalizability by ensuring exposure to diverse secure coding patterns rather than specific fix implementations.

Our final dataset contains 378 manually validated code snippets, balanced across vulnerable and non-vulnerable samples for each language (\autoref{tab:num_snippet_per_lang_merged}). It includes Python (38/38), Java (42/42), C++ (36/36), C (44/44), and JavaScript (29/29). Most samples were drawn from the CWE-snippets and CVEFixes datasets, while JVD provided the non-vulnerable JavaScript cases. To account for the hierarchical nature of CWE, predictions matching a child CWE (e.g., CWE-78) for a snippet labeled with its parent (e.g., CWE-77) were considered correct, with the parent assigned as the final class.

\begin{table}[t]
\centering
\small
\caption{Number of code snippets (vul. / non-vul.).}
\label{tab:num_snippet_per_lang_merged}
\renewcommand{\arraystretch}{1}
\setlength{\tabcolsep}{7.4pt}
\begin{tabular}{lcccc}
\toprule
\textbf{Language} & \textbf{CVE fixes} & \textbf{CWE snippets} & \textbf{JSV} & \textbf{Total} \\
\midrule
C          & 15 / 7   & 29 / 37 & 0 / 0  & 44 / 44 \\
C++        & 12 / 15  & 24 / 21 & 0 / 0  & 36 / 36 \\
Java       & 16 / 3   & 26 / 39 & 0 / 0  & 42 / 42 \\
JavaScript & 15 / 4   & 14 / 2  & 0 / 23 & 29 / 29 \\
Python     & 22 / 38  & 16 / 0  & 0 / 0  & 38 / 38 \\
\midrule
\textbf{Total} & 80 / 67 & 109 / 99 & 0 / 23 & 189 / 189 \\
\bottomrule
\end{tabular}
\end{table}


\begin{table}[t]
\scriptsize
\def\arraystretch{1.1}%
\setlength{\tabcolsep}{5pt}
\caption{LLMs evaluated in this work.}
\centering
\begin{tabularx}{\columnwidth}{@{}l*{4}{>{\centering\arraybackslash}X}@{}}
\toprule
\textbf{Model} & \textbf{Parameters} & \textbf{Architecture} & \textbf{Input Tokens} & \textbf{Output Tokens} \\
\midrule
GPT-3.5-Turbo~\cite{gpt35turbo2024} & $>$ 100B & Decoder & 16,385 & 4,096 \\
GPT-4 Turbo~\cite{gpt4turbo2024} & $>$ 100B & Decoder & 128K & 4,096 \\
GPT-4o~\cite{GPT4o2024} & $>$ 100B & Decoder & 128K & 4,096 \\
CodeLLama-7B~\cite{rozière2024codellamaopenfoundation} & 7B & Decoder & 100K & 4,096 \\
CodeLLama-13B~\cite{rozière2024codellamaopenfoundation} & 13B & Decoder & 100K & 4,096 \\
Gemini 1.5 Pro~\cite{geminiteam2024gemini15unlockingmultimodal} & $>$ 100B & Decoder & 1M & 8,192 \\
\bottomrule
\end{tabularx}
\label{tab:llm-list}
\end{table}


\begin{table*}[!htbp]
\scriptsize
\def\arraystretch{1.3}
\setlength{\tabcolsep}{6pt}
\caption{System and user prompts: Sys1, VD-Sys2, VD-User1 and VD-User2 are used for vulnerability detection (RQ\textsubscript{1}). CWE-Sys1 and CWE-Sys2 are used for system prompts 1 and 2 for CWE Classification. CWE-UserZ (zero-shot) and CWE-UserF (few-shot) are user prompts used for CWE classification (RQ\textsubscript{2}).}
\centering
\begin{tabularx}{\textwidth}{@{}l>{\raggedright\arraybackslash}p{2.2cm}>{\raggedright\arraybackslash}X>{\raggedright\arraybackslash}p{2cm}@{}}
\toprule
\textbf{No.} & \textbf{Prompt Type} & \textbf{Prompt Template} & \textbf{Verbalizer}\\
\midrule
V-Sys1 & System Description & You are an AI binary vulnerability classifier that identifies whether the provided code is vulnerable or not vulnerable. You should respond with either only 'vulnerable' or 'not vulnerable'. & N/A \\ \hline

V-Sys2 & System Description & You are an experienced developer who knows the security vulnerability very well. & N/A\\ \hline

U1 & User Description for VD & Classify the following code in vulnerable or not vulnerable. Output either only 'vulnerable' or 'not vulnerable'.& vulnerable\newline or \newline non-vulnerable \\ \hline

U2 & User Description for VD & Now you need to identify whether a code contains a vulnerability or not. If has any potential vulnerability, output: 'vulnerable'. Otherwise, output: 'not vulnerable'. \newline You must respond with either 'vulnerable' or 'not vulnerable' only. The code is below.& vulnerable\newline or \newline non-vulnerable \\ \hline

C-Sys1 & System Description for CWE & You are an AI vulnerability classifier that identifies CWE of a sent code, if that code has a vulnerability. \newline You should respond with only one CWE. You don't have to respond with any other information except for one of the CWEs. & N/A\\ \hline

C-Sys2 & System Description for CWE Classification & You are an experienced developer who knows the security vulnerability very well. & N/A\\ \hline

UZ & User Description for CWE (zero-shot) & Classify the following code in CWE a category. Output only one CWE tag in lowercase letters. Do not provide the full name of the CWE. Respond with 'non-vul' if you think the code is not vulnerable. & CWE ID \\\hline

UF & User Description for CWE Classification with the few-shot setting & Classify the following code in a CWE category. The code should have a vulnerability that corresponds to one of the top-25 CWEs below. Output only one CWE tag in lowercase letters. Do not provide the full name of the CWE. Respond 'non-vul' if you think the code is not vulnerable. Below is the list of the top 25 CWEs: \newline CWE-787: Out-of-bounds Write \newline ... \newline CWE-276: Incorrect Default Permissions. & CWE ID \\
\bottomrule
\end{tabularx}
\label{table:prompts_design}
\end{table*}

\subsection{Large Language Models} 

We evaluate the effectiveness of using six state-of-the-art pre-trained LLMs.
\autoref{tab:llm-list} shows the selected LLMs, including their parameter size (the number of parameters for GPT and Gemini 1.5 Pro models are not disclosed by the organizations, and therefore, we report an estimate), architecture, input, and output token sizes. 
Specifically, we considered GPT-3.5 Turbo, GPT-4 Turbo, and CodeLlama models available as of June 2024, and pertaining from different organizations such as OpenAI's GPT-3.5-Turbo~\cite{gpt35turbo2024}, GPT-4 Turbo~\cite{gpt4turbo2024} and GPT-4o~\cite{GPT4o2024}, as well as Meta's CodeLLama-7B~\cite{rozière2024codellamaopenfoundation} and CodeLLama-13B~\cite{rozière2024codellamaopenfoundation}, and Google's Gemini 1.5 Pro~\cite{geminiteam2024gemini15unlockingmultimodal}. 

For the GPT models, we utilized the supporting company's Azure OpenAI Studio. Concerning CodeLLama-7B and CodeLLama-13B models, we employed Google Colab and the \texttt{transformers} library by HuggingFace~\cite{wolf-etal-2020-transformers}. For the Gemini 1.5 Pro model, we used the Vertex AI from the Google Cloud Platform. Lastly, to minimize the models' creativity and obtain more focused, conservative, and consistent responses, we set the temperature parameter to $0.1$ for all LLMs, as done in previous studies~\cite{chen2023chatgptsbehaviorchangingtime}.

\begin{table}[t]
    \centering
    \scriptsize
    \def\arraystretch{1}%
    \setlength{\tabcolsep}{2.5pt}
    \caption{RQ\textsubscript{1}: Effectiveness comparison of six LLMs using four prompt configurations for CWE detection.}
    \label{table:combined_comparison_vul}
    \begin{tabular}{lcccccccccccc}
    
    \toprule
    
    \multicolumn{1}{c}{\multirow{2}{*}{\textbf{LLM / Prompt}}} 
    & \multicolumn{3}{c}{\textbf{V-Sys1–U1}} 
    & \multicolumn{3}{c}{\textbf{V-Sys1–U2}} 
    & \multicolumn{3}{c}{\textbf{C-Sys2–U1}} 
    & \multicolumn{3}{c}{\textbf{C-Sys2–U2}} \\
    \cmidrule(lr){2-4}\cmidrule(lr){5-7}\cmidrule(lr){8-10}\cmidrule(lr){11-13}
      & Prec. & Rec. & $F_1$ & Prec. & Rec. & $F_1$ & Prec. & Rec. & $F_1$ & Prec. & Rec. & $F_1$ \\
    \midrule

    \sc All \\    
        \quad GPT-3.5 Turbo & 0.59 & 0.77 & 0.67 & 0.55 & 0.82 & 0.66 & 0.73 & 0.70 & 0.71 & 0.58 & 0.77 & 0.66\\
        \quad GPT-4 Turbo & \textbf{0.81} & 0.82 & 0.81 & 0.73 & 0.92 & 0.81 & 0.75 & 0.87 & 0.81 & 0.71 & 0.93 & 0.81\\
        \quad GPT-4o & 0.72 & 0.97 & \textbf{0.83} & 0.64 & 0.99 & 0.78 & 0.57 & \textbf{1.00} & 0.72 & 0.57 & 0.99 & 0.73\\
        \quad CodeLlama-7b & 0.58 & 0.89 & 0.70 & 0.54 & 0.93 & 0.68 & 0.52 & 0.98 & 0.68 & 0.55 & 0.84 & 0.66\\
        \quad CodeLlama-13b & 0.76 & 0.28 & 0.41 & 0.67 & 0.65 & 0.66 & 0.62 & 0.77 & 0.69 & 0.62 & 0.81 & 0.71\\
        \quad Gemini 1.5 Pro & 0.79 & 0.83 & 0.81 & 0.65 & 0.84 & 0.73 & 0.71 & 0.83 & 0.77 & 0.62 & 0.97 & 0.75\\[0.5em]
    
    \sc Python \\
        \quad GPT-3.5 Turbo & 0.59 & 0.87 & 0.70 & 0.54 & 0.89 & 0.67 & 0.75 & 0.71 & 0.73 & 0.57 & 0.87 & 0.69 \\
        \quad GPT-4 Turbo & 0.73 & 0.79 & 0.76 & 0.72 & 0.89 & \textbf{0.80} & 0.73 & 0.87 & \textbf{0.80} & 0.66 & 0.87 & 0.75 \\
        \quad GPT-4o & 0.70 & 0.92 & \textbf{0.80} & 0.60 & 0.95 & 0.73 & 0.54 & \textbf{1.00} & 0.70 & 0.55 & 0.97 & 0.70 \\
        \quad CodeLlama-7b & 0.67 & 0.76 & 0.72 & 0.56 & 0.79 & 0.65 & 0.52 & 0.97 & 0.68 & 0.55 & 0.76 & 0.64 \\
        \quad CodeLlama-13b & 0.69 & 0.24 & 0.35 & 0.59 & 0.61 & 0.60 & 0.59 & 0.63 & 0.61 & 0.53 & 0.63 & 0.58 \\
        \quad Gemini 1.5 Pro & \textbf{0.79} & 0.71 & 0.75 & 0.59 & 0.71 & 0.64 & 0.70 & 0.84 & 0.76 & 0.60 & 0.95 & 0.73 \\[0.5em]
    
    \sc C \\ 
        \quad GPT-3.5 Turbo & 0.55 & 0.68 & 0.61 & 0.48 & 0.70 & 0.57 & 0.68 & 0.73 & 0.70 & 0.56 & 0.68 & 0.61 \\
        \quad GPT-4 Turbo & 0.78 & 0.91 & 0.84 & 0.67 & 0.98 & 0.80 & 0.75 & 0.89 & 0.81 & 0.70 & 0.98 & 0.82 \\
        \quad GPT-4o & 0.75 & \textbf{1.00} & \textbf{0.85} & 0.63 & \textbf{1.00} & 0.77 & 0.58 & \textbf{1.00} & 0.73 & 0.59 & \textbf{1.00} & 0.75 \\
        \quad CodeLlama-7b & 0.55 & 0.93 & 0.69 & 0.55 & 0.98 & 0.70 & 0.51 & 0.95 & 0.67 & 0.51 & 0.86 & 0.64 \\
        \quad CodeLlama-13b & \textbf{0.93} & 0.32 & 0.47 & 0.67 & 0.77 & 0.72 & 0.60 & 0.86 & 0.71 & 0.72 & 0.89 & 0.80 \\
        \quad Gemini 1.5 Pro & 0.73 & 0.86 & 0.79 & 0.67 & 0.80 & 0.73 & 0.79 & 0.77 & 0.78 & 0.65 & \textbf{1.00} & 0.79 \\[0.5em]
    
    \sc C++ \\
        \quad GPT-3.5 Turbo & 0.57 & 0.75 & 0.65 & 0.56 & 0.86 & 0.68 & 0.71 & 0.75 & 0.73 & 0.59 & 0.75 & 0.66 \\
        \quad GPT-4 Turbo & 0.82 & 0.89 & 0.85 & 0.76 & 0.97 & 0.85 & 0.71 & 0.94 & 0.81 & 0.73 & 0.97 & 0.83 \\
        \quad GPT-4o & 0.78 & \textbf{1.00} & \textbf{0.88} & 0.67 & \textbf{1.00} & 0.80 & 0.57 & \textbf{1.00} & 0.73 & 0.55 & \textbf{1.00} & 0.71 \\
        \quad CodeLlama-7b & 0.57 & 0.92 & 0.70 & 0.54 & 0.94 & 0.69 & 0.54 & \textbf{1.00} & 0.70 & 0.56 & 0.94 & 0.70 \\
        \quad CodeLlama-13b & 0.83 & 0.28 & 0.42 & 0.63 & 0.61 & 0.62 & 0.64 & 0.83 & 0.72 & 0.64 & 0.89 & 0.74 \\
        \quad Gemini 1.5 Pro & \textbf{0.85} & 0.92 & \textbf{0.88} & 0.67 & 0.89 & 0.76 & 0.67 & 0.89 & 0.76 & 0.61 & 0.97 & 0.75 \\[0.5em]
    
    \sc JavaScript \\
        \quad GPT-3.5 Turbo & 0.69 & 0.86 & 0.77 & 0.63 & 0.93 & 0.75 & 0.85 & 0.76 & 0.80 & 0.67 & 0.90 & 0.76 \\
        \quad GPT-4 Turbo & \textbf{0.88} & 0.79 & 0.84 & 0.77 & 0.93 & 0.84 & 0.74 & 0.86 & 0.79 & 0.67 & 0.90 & 0.76 \\
        \quad GPT-4o & 0.61 & 0.97 & 0.75 & 0.59 & \textbf{1.00} & 0.74 & 0.52 & \textbf{1.00} & 0.68 & 0.52 & 0.97 & 0.67 \\
        \quad CodeLlama-7b & 0.58 & 0.97 & 0.73 & 0.53 & 0.97 & 0.68 & 0.51 & \textbf{1.00} & 0.67 & 0.60 & 0.90 & 0.72 \\
        \quad CodeLlama-13b & 0.57 & 0.28 & 0.37 & 0.72 & 0.62 & 0.67 & 0.64 & 0.72 & 0.68 & 0.58 & 0.86 & 0.69 \\
        \quad Gemini 1.5 Pro & 0.87 & 0.90 & \textbf{0.88} & 0.60 & 0.86 & 0.70 & 0.67 & 0.83 & 0.74 & 0.56 & \textbf{1.00} & 0.72 \\ [0.5em]
        
    \sc Java \\
        \quad GPT-3.5 Turbo & 0.57 & 0.74 & 0.65 & 0.56 & 0.76 & 0.65 & 0.69 & 0.57 & 0.62 & 0.57 & 0.69 & 0.62 \\
        \quad GPT-4 Turbo & \textbf{0.86} & 0.71 & 0.78 & 0.76 & 0.83 & 0.80 & 0.83 & 0.81 & 0.82 & 0.78 & 0.93 & \textbf{0.85} \\
        \quad GPT-4o & 0.75 & 0.98 & \textbf{0.85} & 0.70 & \textbf{1.00} & 0.82 & 0.61 & \textbf{1.00} & 0.76 & 0.64 & \textbf{1.00} & 0.78 \\
        \quad CodeLlama-7b & 0.54 & 0.88 & 0.67 & 0.52 & 0.95 & 0.67 & 0.52 & \textbf{1.00} & 0.68 & 0.54 & 0.76 & 0.63 \\
        \quad CodeLlama-13b & 0.75 & 0.29 & 0.41 & 0.78 & 0.60 & 0.68 & 0.65 & 0.79 & 0.71 & 0.62 & 0.79 & 0.69 \\
        \quad Gemini 1.5 Pro & 0.74 & 0.76 & 0.75 & 0.70 & 0.93 & 0.80 & 0.74 & 0.83 & 0.79 & 0.65 & 0.95 & 0.77 \\ 
    
    \bottomrule
\end{tabular}
\end{table}

\subsection{Classification Methods}

Our study examined three classification approaches.

\head{Vulnerability Detection}
A simple binary classification method in which the LLM is tasked to categorize an input snippet into either ``vulnerable'' or ``not vulnerable''. 

\head{CWE classification (zero-shot)} 
A multi-class classification method in which the LLM is tasked to categorize an input snippet as pertaining to a specific CWE. The zero-shot setting provides only a task description in natural language as part of the input prompt.

\head{CWE classification (few-shot)} 
Same multi-class classification method as before, but using a prompt that provides the top-25 CWE labels~\cite{MITRE2023CWE} to elicit CWE-specific reasoning and map the input snippet to the corresponding CWE category.

\subsection{Prompt and Input Data}\label{sec:prompts}

LLM's prompts were divided into two main types: \textit{system prompt}, which defines the behavior, and \textit{user prompt}, which gives instructions regarding a given task. 

In our study, the final prompts combine one system prompt with one user prompt. 
\autoref{table:prompts_design} reports all the prompts that were used for our experiments. We considered two system prompts: VD-Sys1 gives the LLM a \textit{detailed} role for vulnerability detection. In contrast, VD-Sys2 had a more \textit{generic} role, as described by Zhou et al.~\cite{zhou2024}. Similarly, we designed two different types of user prompts to detect a vulnerability: U1 presents a concise request for vulnerability detection, whereas VD-User2 provides more descriptive instruction as used by Zhou et al.~\cite{zhou2024}.

For the classification of CWE (RQ\textsubscript{2}), we introduced two system prompts (CWE-Sys1 and CWE-Sys2) and two user prompts (CWE-UserZ and CWE-UserF). 
CWE-Sys1 is a similar prompt compared to VD-Sys1, which provides a more specific role for CWE classification. Conversely, CWE-Sys2 is the same prompt as VD-Sys2~\cite{zhou2024}. 
CWE-UserZ is a user prompt for CWE classification with a zero-shot setting (i.e., without any examples). CWE-UserF is for CWE classification experiments with the few-shot setting, and it has a list of the top 25 CWEs comparable to CWE-UserZ. To form the final prompt text, we concatenated an input code snippet with one of our described prompt strategies. 

\subsection{Procedure and Metrics}

To answer RQ\textsubscript{1} and RQ\textsubscript{2}, we executed each LLM on all code snippets of our datasets, using the prompt strategies described in \autoref{sec:prompts}. As evaluation metrics, we employed four metrics, namely accuracy, recall, precision, and $F_1$ score, commonly used to evaluate the performance of LLMs in vulnerability detection. 
In our evaluation, valid detections/classifications were treated as the positive class, while detection/classification augmentations were considered the negative class. We accounted for CWE hierarchical relationships, considering positive scores for cases where the LLM predicted a code snippet labeled with a specific CWE (e.g., CWE-77) as one of its child CWEs (e.g., CWE-78), according to the MITRE CWE database~\cite{mitre}.
For CWE classification (RQ\textsubscript{2}), we evaluated the overall performance by macro-averaging the precision parameter across all classes in the precision, recall, and $F_1$ score. 

Overall, in our evaluation, we computed 18,144 predictions (6 LLMs $\times$ 378 code snippets $\times$ 8 prompt configurations), which accounted for 10 hours of computing time.


\begin{table}[t]
    \centering
    \scriptsize
    \def\arraystretch{1}%
    \setlength{\tabcolsep}{2.5pt}
    \caption{RQ\textsubscript{2}: Effectiveness comparison of six LLMs using four prompt configurations for CWE classification (Precision, Recall, $F_1$).}
    \label{table:combined_comparison}
    \begin{tabular}{lcccccccccccc}
    
    \toprule
    
    \multicolumn{1}{c}{\multirow{3}{*}{\textbf{LLM / Prompt}}} & \multicolumn{3}{c}{\textbf{C-Sys1 + UZ}} & \multicolumn{3}{c}{\textbf{C-Sys2 + UZ}} & \multicolumn{3}{c}{\textbf{C-Sys1 + UF}} & \multicolumn{3}{c}{\textbf{C-Sys2 + UF}}
    \\\cmidrule(lr){2-4}\cmidrule(lr){5-7}\cmidrule(lr){8-10}\cmidrule(lr){11-13}
               & Prec. & Rec. & $F_1$ & Prec. & Rec. & $F_1$ & Prec. & Rec. & $F_1$ & Prec. & Rec. & $F_1$\\
               \midrule
    \sc All \\    
    \quad GPT-3.5 Turbo & 0.14 & 0.12 & 0.12 & 0.14 & 0.12 & 0.12 & 0.34 & 0.35 & 0.32 & 0.24 & 0.23 & 0.21\\
    \quad GPT-4 Turbo & 0.21 & 0.18 & 0.18 & 0.30 & 0.23 & 0.25 & 0.42 & 0.52 & 0.44 & 0.54 & 0.58 & 0.53\\
    \quad GPT-4o & 0.23 & 0.19 & 0.20 & 0.22 & 0.19 & 0.19 & \textbf{0.61} & 0.69 & 0.61 & 0.60 & \textbf{0.74} & \textbf{0.63}\\
    \quad CodeLlama-7b & 0.12 & 0.09 & 0.08 & 0.08 & 0.05 & 0.06 & 0.13 & 0.10 & 0.08 & 0.11 & 0.08 & 0.06\\
    \quad CodeLlama-13b & 0.09 & 0.08 & 0.08 & 0.08 & 0.05 & 0.05 & 0.09 & 0.07 & 0.07 & 0.10 & 0.09 & 0.07\\
    \quad Gemini 1.5 Pro & 0.15 & 0.13 & 0.13 & 0.16 & 0.15 & 0.14 & 0.52 & 0.56 & 0.51 & 0.56 & 0.52 & 0.49 \\[0.5em]
    
    \sc Python \\
    \quad GPT-3.5 Turbo & 0.25 & 0.21 & 0.22 & 0.22 & 0.20 & 0.20 & 0.27 & 0.32 & 0.28 & 0.32 & 0.30 & 0.29 \\
    \quad GPT-4 Turbo & 0.43 & 0.42 & 0.40 & 0.42 & 0.35 & 0.36 & 0.53 & 0.54 & 0.48 & 0.59 & 0.55 & 0.52 \\
    \quad GPT-4o & 0.36 & 0.26 & 0.27 & 0.35 & 0.31 & 0.30 & 0.58 & 0.52 & 0.51 & 0.60 & \textbf{0.65} & 0.60 \\
    \quad CodeLlama-7b & 0.09 & 0.07 & 0.07 & 0.16 & 0.10 & 0.11 & 0.00 & 0.02 & 0.01 & 0.15 & 0.14 & 0.11 \\
    \quad CodeLlama-13b & 0.16 & 0.15 & 0.14 & 0.10 & 0.09 & 0.08 & 0.17 & 0.13 & 0.11 & 0.17 & 0.18 & 0.12 \\
    \quad Gemini 1.5 Pro & 0.24 & 0.21 & 0.20 & 0.25 & 0.22 & 0.22 & 0.61 & 0.64 & 0.58 & \textbf{0.72} & 0.62 & \textbf{0.62} \\[0.5em]
    
    \sc C \\ 
    \quad GPT-3.5 Turbo & 0.21 & 0.20 & 0.19 & 0.20 & 0.19 & 0.19 & 0.46 & 0.43 & 0.39 & 0.32 & 0.33 & 0.31 \\
    \quad GPT-4 Turbo & 0.38 & 0.37 & 0.36 & 0.46 & 0.42 & 0.43 & 0.56 & 0.66 & 0.58 & \textbf{0.67} & 0.69 & \textbf{0.66} \\
    \quad GPT-4o & 0.37 & 0.31 & 0.32 & 0.33 & 0.28 & 0.29 & 0.64 & 0.74 & 0.65 & 0.62 & \textbf{0.76} & 0.64 \\
    \quad CodeLlama-7b & 0.19 & 0.15 & 0.15 & 0.08 & 0.06 & 0.07 & 0.12 & 0.20 & 0.13 & 0.12 & 0.17 & 0.12 \\
    \quad CodeLlama-13b & 0.18 & 0.15 & 0.15 & 0.15 & 0.12 & 0.12 & 0.14 & 0.13 & 0.12 & 0.20 & 0.21 & 0.18 \\
    \quad Gemini 1.5 Pro & 0.22 & 0.17 & 0.18 & 0.29 & 0.26 & 0.26 & 0.57 & 0.60 & 0.56 & \textbf{0.67} & 0.62 & 0.61 \\[0.5em]
    
    \sc C++ \\
    \quad GPT-3.5 Turbo & 0.21 & 0.19 & 0.19 & 0.25 & 0.22 & 0.22 & 0.43 & 0.46 & 0.41 & 0.32 & 0.32 & 0.29 \\
    \quad GPT-4 Turbo & 0.39 & 0.39 & 0.37 & 0.44 & 0.42 & 0.41 & 0.59 & 0.68 & 0.60 & 0.57 & 0.66 & 0.57 \\
    \quad GPT-4o & 0.42 & 0.39 & 0.39 & 0.40 & 0.38 & 0.36 & \textbf{0.61} & 0.75 & \textbf{0.64} & 0.59 & \textbf{0.76} & 0.62 \\
    \quad CodeLlama-7b & 0.17 & 0.15 & 0.14 & 0.18 & 0.12 & 0.14 & 0.17 & 0.18 & 0.16 & 0.09 & 0.13 & 0.09 \\
    \quad CodeLlama-13b & 0.18 & 0.15 & 0.15 & 0.12 & 0.11 & 0.10 & 0.22 & 0.20 & 0.19 & 0.27 & 0.22 & 0.19 \\
    \quad Gemini 1.5 Pro & 0.32 & 0.28 & 0.29 & 0.32 & 0.30 & 0.30 & 0.45 & 0.40 & 0.39 & 0.45 & 0.42 & 0.41 \\[0.5em]
        
    \sc JavaScript \\
    \quad GPT-3.5 Turbo & 0.25 & 0.26 & 0.24 & 0.27 & 0.27 & 0.25 & 0.42 & 0.50 & 0.44 & 0.41 & 0.40 & 0.37 \\
    \quad GPT-4 Turbo & 0.28 & 0.28 & 0.26 & 0.52 & 0.52 & 0.50 & 0.52 & 0.64 & 0.53 & \textbf{0.68} & 0.71 & \textbf{0.66} \\
    \quad GPT-4o & 0.37 & 0.42 & 0.38 & 0.29 & 0.29 & 0.27 & 0.59 & 0.62 & 0.59 & 0.53 & 0.61 & 0.55 \\
    \quad CodeLlama-7b & 0.16 & 0.20 & 0.17 & 0.25 & 0.25 & 0.22 & 0.13 & 0.19 & 0.13 & 0.05 & 0.12 & 0.06 \\
    \quad CodeLlama-13b & 0.26 & 0.30 & 0.26 & 0.27 & 0.27 & 0.23 & 0.08 & 0.07 & 0.06 & 0.13 & 0.11 & 0.09 \\
    \quad Gemini 1.5 Pro & 0.31 & 0.35 & 0.33 & 0.27 & 0.28 & 0.26 & 0.60 & \textbf{0.73} & 0.63 & 0.52 & 0.52 & 0.49 \\[0.5em]
    
    \sc Java \\
    \quad GPT-3.5 Turbo & 0.19 & 0.16 & 0.16 & 0.22 & 0.18 & 0.19 & 0.28 & 0.31 & 0.29 & 0.22 & 0.23 & 0.21 \\
    \quad GPT-4 Turbo & 0.29 & 0.23 & 0.25 & 0.42 & 0.35 & 0.37 & 0.49 & 0.55 & 0.50 & 0.48 & 0.52 & 0.48 \\
    \quad GPT-4o & 0.35 & 0.30 & 0.31 & 0.36 & 0.32 & 0.32 & 0.59 & 0.63 & 0.59 & \textbf{0.63} & 0.68 & \textbf{0.64} \\
    \quad CodeLlama-7b & 0.14 & 0.10 & 0.10 & 0.11 & 0.11 & 0.10 & 0.07 & 0.11 & 0.07 & 0.09 & 0.13 & 0.08 \\
    \quad CodeLlama-13b & 0.17 & 0.18 & 0.17 & 0.16 & 0.16 & 0.14 & 0.14 & 0.15 & 0.12 & 0.15 & 0.12 & 0.12 \\
    \quad Gemini 1.5 Pro & 0.28 & 0.26 & 0.26 & 0.29 & 0.26 & 0.26 & 0.59 & \textbf{0.69} & 0.59 & 0.52 & 0.60 & 0.52 \\
    
    \bottomrule
\end{tabular}
\end{table}

\section{Experimental Results}\label{sec:study}

\subsection{RQ\textsubscript{1} (Vulnerability detection)}

\autoref{table:combined_comparison_vul} (vulnerability detection) presents the effectiveness results across the different prompt setups.
Overall, GPT-4 Turbo and GPT-4o showed the best binary classification scores. The former achieved the highest accuracy and precision (0.81), whereas the latter showed the highest recall (1.00) and $F_1$ score (0.83). 
%
%
Changing the prompt from Sys1 to VD-Sys2 (i.e., a more general role definition) had a positive impact only for GPT-3.5 Turbo and CodeLlama-13b, proving detrimental for the other LLMs. Although VD-Sys2 configurations achieved the highest recall results, they only marginally outperformed Sys1 configurations and came at the cost of reduced precision. The best balance between detecting vulnerabilities and minimizing false positives was achieved by the prompt configurations Sys1–U1 and Sys1 + VD-User2 (the first two macro-columns). Among these, Sys1–U1 was the most effective for minimizing false alarms (GPT-4 Turbo) and detecting vulnerabilities (GPT-4o).

For Python, GPT-4 Turbo and GPT-4o consistently outperformed other models. In particular, both models achieved the highest $F_1$ score of 0.80 using the U1 task prompt. Although Gemini 1.5 pro delivered the highest accuracy, its $F_1$ score did not reach more than 0.80 due to its lower recall than GPT-4 Turbo and GPT-4o. 

In C, GPT-4o demonstrated exceptional recall, achieving perfect scores (1.00) across all prompt configurations. Although the model size of CodeLlama-13b is considerably smaller than other bigger models, such as GPT-4o and GPT-4 Turbo, it achieved the highest precision of 0.93. This revealed the potential to reliably identify true positive vulnerabilities with a smaller sized LLM.

For C++, both GPT-4 Turbo and Gemini 1.5 Pro showed a solid performance, both achieving the highest $F_1$ score of 0.88 using Sys1–U1. Particularly, GPT-4o surpassed other models in recall (1.0) across all prompt sets, which is similar to C. 

In JavaScript, Gemini 1.5 Pro maintained a balance between precision (0.87) and recall (0.90) in Sys1–U1; thus, it achieved the highest $F_1$ score (0.88). Although GPT-4o obtained the highest recall of 1.0 using two prompt sets, CodeLlama-7b, a model more than 10 times smaller in size, also reached a recall of 1.0 (VD-Sys2 + U1). 

Lastly, regarding the result in Java, GPT-4 Turbo and GPT-4o showed the best $F_1$ score of 0.85 using Sys1–U1 used by GPt-4o and VD-Sys2 + VD-User2 used by GPT-4 Turbo. While other models did not achieved satisfactory performance, GPT-4 Turbo exhibited the highest precision across all prompt sets. In particular, Sys1–U1, the highest precision of 0.86 (GPT-4 Turbo), was approximately 15\% better than the second highest one (GPT-4o). \\

\rqbox{
\textbf{RQ\textsubscript{1} (vulnerability detection)}: \textit{
Overall, GPT-4 Turbo and GPT-4o achieved the best vulnerability detection scores across multiple languages ($F_1=0.81$ and $F_1=0.83$). GPT-4 Turbo was the most effective for minimizing false alarms. GPT-4 Turbo achieved the highest $F_1$ score in Python. GPT-4o demonstrated exceptional recall, particularly in C and C++. Gemini 1.5 Pro performed the best in JavaScript. GPT-4-based models were most effective in Java.
}
}

\subsection{RQ\textsubscript{2} (CWE classification)}

\autoref{table:combined_comparison} (CWE classification) presents the effectiveness results across the different prompt setups. Unlike the detection task, the scores vary significantly across configurations. Similarly to the detection task, overall, GPT-4 Turbo and GPT-4o showed the best multi-class classification scores. Classification demonstrates a more difficult task than classification, as GPT-4 Turbo achieved an $F_1$ score of $0.53$ and GPT-4o achieved an $F_1$ score of $0.63$. The best results are obtained using the few-shot learning technique and a generic role (CWE-Sys2 + CWE-UserF configuration). 

The effectiveness of GPT-3.5 Turbo, GPT-4 Turbo, GPT-4o, and Gemini 1.5 Pro, improved using the few-shot learning technique. GPT-4o's $F_1$ score improves by approximately tripling in all prompt sets when switching from zero-shot to few-shot. Similarly, GPT-4 Turbo shows substantial improvements with more than double the recall and $F_1$ score between CWE-Sys2 + CWE-UserZ and CWE-Sys2 + CWE-UserF.

For Python, Gemini 1.5 Pro was the most effective model using CWE-Sys2 + CWE-UserF setting. This exhibited the highest accuracy ($>$ 0.70) and $F_1$ (0.62). 

In C, GPT-4 Turbo obtained balanced precision and recall (0.67 and 0.69), which resulted in the highest $F_1$ score among all six LLMs using the CWE-Sys2 + CWE-UserF prompt set. Moreover, similar to the result in RQ\textsubscript{1}, GPT-4o achieved a remarkable recall of 0.76. 

For C++, GPT-4o consistently outperformed other models across precision, recall, and $F_1$ score, achieving the highest $F_1$ score (0.64) and recall (0.76) with both prompts for the few-shot setting. 

Regarding JavaScript, GPT-4 Turbo was the most effective LLM using the CWE-Sys2 + CWE-UserF prompt configuration, especially in accuracy, precision, and $F_1$ score. 

Lastly, GPT-4o demonstrated the best overall effectiveness, with the highest $F_1$ score (0.64) and precision (0.63) using CWE-Sys2 + CWE-UserF. Across all languages, the few-shot learning approach (CWE-UserF) consistently yielded better results compared to zero-shot prompts. \\

\rqbox{
\textbf{RQ\textsubscript{2} (CWE classification)}: \textit{
GPT-4o achieved the best overall vulnerability classification performance across languages (\(recall = 0.74\), \(F_{1} = 0.63\)). 
GPT-4 Turbo performed best in C and JavaScript, while Gemini~1.5~Pro outperformed in Python. 
Across all models and languages, few-shot prompting consistently outperformed zero-shot.
}
}

\section{User Study}\label{sec:userstudy}

\subsection{\codeguardian} \label{sec:deployment}

To help engineers understand potential vulnerabilities in source code within their development workflow, we developed \codeguardian, a VSCode extension that features just-in-time LLM-powered vulnerability scanning. It supports developers by providing real-time vulnerability analysis of their source code. More specifically, a user can highlight code snippets in editor, and \codeguardian provides instant vulnerability analysis of the highlighted code. This feature helps developers identify security flaws at a snippet level. Multi-snippet or multi-file analyses are not supported.

Our study identified GPT-4 Turbo and GPT-4o as the best models for CWE detection and classification (see \autoref{sec:study}). Consequently, \codeguardian provides users with the option to select these models, using Azure OpenAI Studio APIs in the back end. The back-end system receives the highlighted code from a user and processes it to be concatenated to a specific prompt. After the vulnerability analysis, the results are shown on the extension panel. 
Additional details of \codeguardian are available in our replication package~\cite{replication-package}. 

The vulnerability analysis in \codeguardian comprises two stages: first, a vulnerability detection phase, to determine whether the code is vulnerable or non-vulnerable, and second, a CWE classification phase, employing a few-shot setting for vulnerable code. Finally, the output of \codeguardian is a detailed analysis of the CWE classification results.

\subsection{Experiment Design}

We evaluated \codeguardian by a dedicated research question, with the main objective of providing a qualitative assessment of LLMs for security vulnerability detection.

\textbf{RQ\textsubscript{3} (usefulness):} Does \codeguardian improve developers' accuracy and efficiency in vulnerability detection?

To address RQ\textsubscript{3}, we performed a user study within an industrial setting, which we describe next.

\head{Participants and Grouping}
Our experiments for the user study involved a diverse group of participants who have industrial software engineering experience. A total of 22 participants were selected, and we asked them to answer our pre-questionnaire. The pre-questionnaire contains several questions asking each participant about their experience with software development, software development skills, knowledge of cybersecurity, and familiarity with five selected programming languages. We used a 5-point Likert scale for the skills and knowledge questions and a nominal scale for the rest of the questions.


\begin{table}[t]
\centering
\def\arraystretch{1.1}%
\setlength{\tabcolsep}{8.5pt}
\caption{An overview of the six challenges of our user study.}
\label{tab:tasks_usability_study}
\begin{tabular}{>{\raggedright\arraybackslash}m{0.4cm} >{\raggedright\arraybackslash}m{1.2cm} >{\raggedright\arraybackslash}m{5.1cm}}
\toprule
\textbf{No} & \textbf{Language} & \textbf{Vulnerability} \\
\midrule
1 & C & CWE-125: Out-of-bounds Read \\
2 & C++ & CWE-78: OS Command Injection \\
3 & Java & CWE-190: Integer Overflow or Wraparound \\
4 & JavaScript & CWE-22: Path Traversal \\
5 & Python & Placebo (Not Vulnerable) \\
6 & Python & CWE-79: Cross-site Scripting \\
\bottomrule
\end{tabular}
\end{table}

Participants were divided into two groups based on their responses: an experimental group (using \codeguardian) and a control group (using only Internet search). Each group consists of 11 participants. By balancing expertise in software development and cybersecurity, we ensured that the experimental and control groups had comparable backgrounds.  



We mainly referred to each participant's years of experience in software development and cybersecurity to allocate them into equally distributed groups. Lastly, none of the participants had any experience with \codeguardian.

\head{Challenges Design}
All participants were required to perform a set of challenges during the experiment, which was identifying possible software vulnerabilities. In this experiment, a total of six challenges across five programming languages were presented to both groups. Table~\ref{tab:tasks_usability_study} shows all the code snippets used for this user study, each snippet contains a specific vulnerability. While five snippets were vulnerable, one of the Python codes was used as a Placebo (i.e., not vulnerable). 

\head{Independent Variable} The independent variable was the type of tool used by the participants: \codeguardian (LLM-based) and Internet search/own knowledge. More specifically, Level 1 expects participants to use \codeguardian to identify vulnerabilities, and Level 2 expects them to use their knowledge.

\head{Dependent variables} The experiment for this user study has two dependent variables: (1) challenge completion time and (2) challenge completion accuracy.

\noindent\textbf{Experimental Procedure.}
The first author of this study demonstrated how to use \codeguardian exclusively for the experimental group, taking approximately 2-3 minutes to explain the usage instructions. Each participant then evaluated six code snippets, identifying any vulnerabilities and naming them or providing their CWE labels. The experimental group used \codeguardian to address these challenges, while the control group could only use Internet resources, \textit{except for any LLM-based application}. Participants had a maximum of 5 minutes to answer each question; if they finished early, the session progressed to the next task. 

All participants worked on a dedicated workstation to ensure consistent environmental conditions for all participants. This workstation was an EC2 instance running Debian 12 OS, accessible to all participants via a browser-based VNC.

\head{Data Analysis}
To compare dependent variables across the experimental group and the control group, we analyzed the results using the Mann-Whitney U test~\cite{mann1947test} (with $\alpha = 0.05$) and the magnitude of the differences using the rank-biserial correlation to calculate the effect size. We selected this non-parametric test as the completion time and accuracy data were not normally distributed.



\subsection{RQ\textsubscript{3} (usefulness)}

\begin{figure}[t]
\centering
\includegraphics[width=0.75\columnwidth]{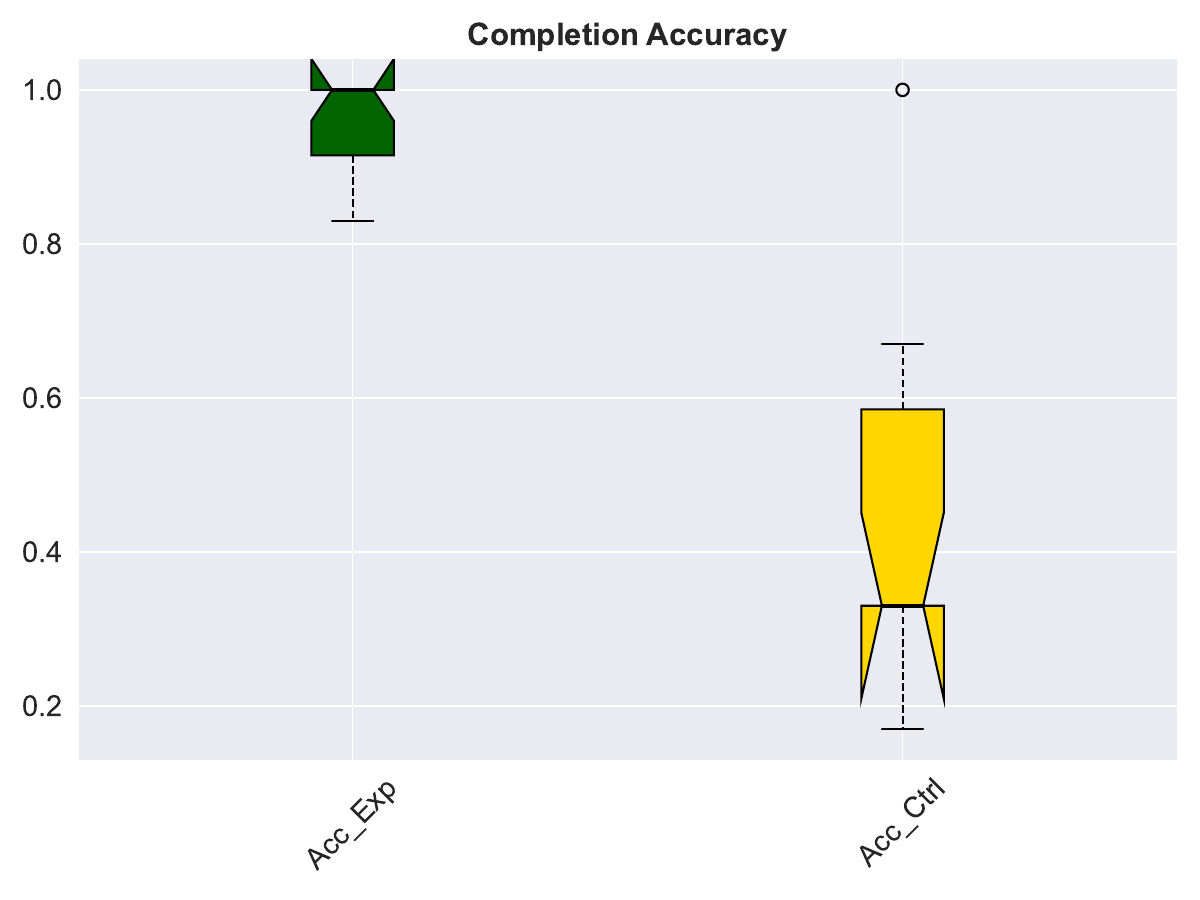}
\caption{Challenge completion accuracy between the experimental group (green) and control group (yellow).}
\label{fig:total_accuracy_boxplot}
\end{figure}

\subsubsection{Completion Accuracy}

Accuracy was calculated based on the amount of correct answers out of all six challenges. \autoref{fig:total_accuracy_boxplot} visualizes the boxplot of challenge completion accuracy for both groups. We used a one-sided research hypothesis since we assumed that the experimental group had a higher percentage of correct answers than the control group. Therefore, the null and alternative hypotheses were defined as follows: $H_0$: The two groups are the same in terms of the total challenge completion accuracy; $H_1$: Total challenge completion accuracy in the experiment group was higher than in the control group.

The experimental group achieved a median accuracy of 1.0 (std = 0.08), while the control group had a median accuracy of 0.33 (std = 0.25). This represents a 203\% increase in accuracy due to the use of \codeguardian. Additionally, the smaller standard deviation in the experimental group's accuracy indicates less variation in correct answer rates among its participants compared to the control group. This suggests that \codeguardian not only enhanced overall accuracy but also contributed to more consistent performance.

The results of the Mann-Whitney U test indicated a statistically significant difference between the two groups (\textit{p}-value=0.0001) with a large effect size (0.88). Thus, we reject the null hypothesis and support the alternative hypothesis, in which the experimental group achieved higher accuracy in completing the challenges. This indicates that the usage of \codeguardian had a strong positive effect on improving the number of correct answers among participants.


\begin{figure}[t]
\centering
\includegraphics[width=0.75\columnwidth]{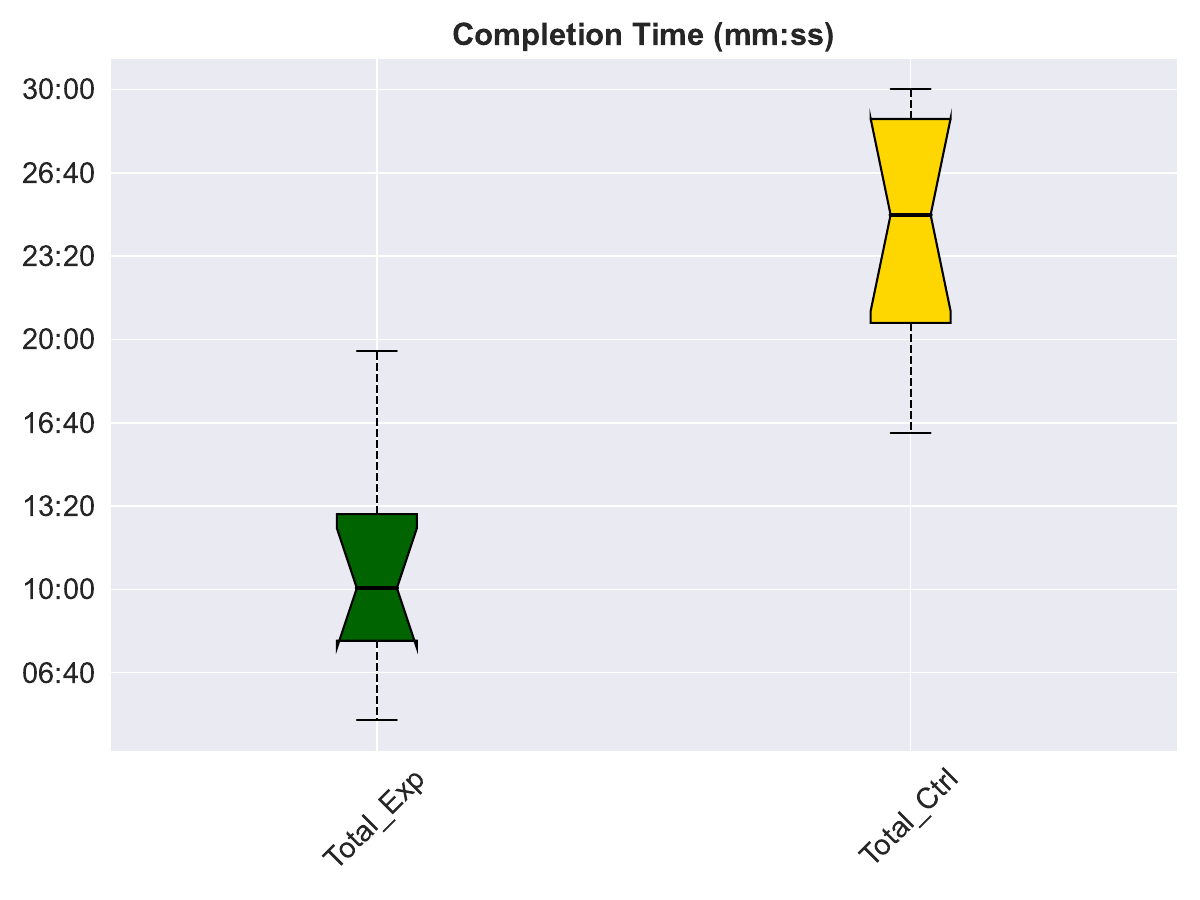}
\caption{Challenge completion time between the experimental group (green) and control group (yellow).}
\label{fig:task_completion_time_boxplot}
\end{figure}

\subsubsection{Completion Time}

We measured the total challenge completion time (minutes:seconds) and compared the results statistically between the experimental and control groups. A one-sided research hypothesis was used for this test, as we hypothesized that the experimental group would complete all the challenges faster than the control group. Hence, the null and alternative hypotheses were defined as follows: $H_0$: The two groups are the same in terms of the total challenge completion time; $H_1$: Total challenge completion time in the experiment group is less than in the control group.

\autoref{fig:task_completion_time_boxplot} depicts the boxplots of completion time over all the challenges. The median and standard deviation in the experimental group were approximately 10:03 (mm:ss), and the standard deviation was 4:29 (mm:ss). On the contrary, the control group exhibited a mean of 24:59 (mm:ss) and a standard deviation of 4:58 (mm:ss). Although the standard deviation of both groups was almost identical, the experimental group completed all the challenges in 66\% less time than the control group on average (around 15 minutes, on average).
The results of the Mann-Whitney U test showed a statistically significant difference (p-value = 0.00018) with a large effect size (0.95). Therefore, we reject the null hypothesis and conclude that the experimental group completed the challenges faster than the control group. \\

\rqbox{
\textbf{RQ\textsubscript{3} (usefulness)}: \textit{\codeguardian significantly improved speed and accuracy, with large effect sizes.}
}
\section{Threats to Validity}\label{sec:ttv}

\head{Internal validity}
All LLMs were compared under identical settings and on the same benchmark. Our testing scripts were validated, but the limited prompt set may affect generalizability. We believe the selected models are representative.
The user study involved 22 participants with prior industrial experience in software development and cybersecurity. Although the sample size was small, it provided sufficient statistical power. Differences in participants' expertise (e.g., programming languages, VSCode) may have influenced results. To mitigate this, we collected background data, balanced expertise across two groups, and ensured identical environments (Debian 12 workstation, 700 Mbps connection). Response times of GPT-4 Turbo occasionally varied due to Azure hosting limitations, which was reflected in post-questionnaire feedback.

\head{External validity}
The evaluation covered a limited set of LLMs and programming languages (C, C++, Python, Java, JavaScript). Results may differ for other models or languages. To reduce this threat, we used state-of-the-art models as of 05.2024. The user study also included a small number of code challenges (one per language plus a Python placebo), constrained by participants' availability. While useful for evaluating \codeguardian, this may not represent the full spectrum of vulnerabilities and real-world scenarios~\cite{Dozono2025LLMSecurityEvaluation}.

 
\section{Discussion}\label{sec:discussion}

\head{Model Selection}
Although GPT-4 Turbo is the predecessor of GPT-4o, it often outperforms other models. While GPT-4o achieved higher recall across all prompt setups, its lower precision caused it to underperform in some cases. This suggests that newer LLM versions do not always surpass earlier ones. Model selection should thus reflect security priorities: GPT-4o is preferable when minimizing false negatives, whereas GPT-4 Turbo better reduces false positives.

\head{Performance Across Languages}
LLM performance varied significantly by programming language. GPT-4 models generally achieved the best results, but Gemini 1.5 Pro showed strong performance in JavaScript (vulnerability detection) and Python (CWE classification). This highlights the importance of considering language-specific performance when using LLMs for secure code review.

\head{Vulnerability Type Analysis}
LLMs were especially effective in detecting injection-related vulnerabilities. Detection rates for CWE-89 (SQL Injection) and CWE-79 (Cross-site Scripting) exceeded 81\% across languages. This strength is particularly relevant for web application security. The user study further confirmed this: in Challenge~6 (CWE-79), participants using \codeguardian completed the task 55\% faster than the control group.

\head{CWE Classification Complexity}
CWE classification proved more difficult than vulnerability detection. Larger models (GPT-4o, GPT-4 Turbo, Gemini 1.5 Pro) clearly outperformed smaller ones, especially with few-shot prompts, which tripled GPT-4o's recall. In contrast, few-shot learning did not improve CodeLlama models, suggesting that smaller models cannot effectively leverage examples due to limited capacity.

\head{User Study}
The experimental group consistently completed tasks faster and with higher accuracy. For instance, in the placebo task (Challenge~5), most participants using \codeguardian answered in under two minutes, compared to nearly five minutes for the control group. Accuracy gains were equally notable: the experimental group achieved over 0.90 on average, versus less than 0.50 in the control group. These results show that \codeguardian accelerates analysis and improves detection, even for developers with limited security expertise. However, strict usage guidelines are needed for junior developers, as LLM outputs remain prone to hallucinations.

\section{Conclusions and Future Work}\label{sec:conclusions}

We presented a comprehensive study on the effectiveness of LLMs for secure code assessment across multiple programming languages. We evaluated six state-of-the-art LLMs on vulnerability detection and CWE classification using a newly developed multi-language dataset. Our results show that GPT-4 Turbo and GPT-4o generally achieved the best overall performance, while Gemini~1.5~Pro performed strongly in JavaScript and Python. 
To bridge research and practice, we introduced \codeguardian, a VSCode extension that integrates LLM-powered vulnerability analysis into developers' workflows. In a user study with 22 industry professionals, \codeguardian improved task completion speed by 66\% and accuracy by 203\% compared to traditional tools.
Future work will focus on expanding and refining the annotated dataset to enhance robustness and generalizability, exploring advanced prompting techniques such as chain-of-thought and retrieval-augmented generation (RAG), other foundation models~\cite{2025-Guo-arxiv,2025-Baresi-ICSE,2025-Maryam-ICST,2026-Merabishvili-ICSEW,2025-Weissl-TOSEM} and tasks such as natural language conversations~\cite{2025-Giebisch-IV,2026-Sorokin-SANER,2025-Habicht-arxiv}, and improving \codeguardian usability through interface enhancements and developer feedback.

\balance
\bibliographystyle{ACM-Reference-Format}
\bibliography{paper}

\end{document}